\documentstyle[12pt]{article}
\begin{document}
\newcommand{\lesssim}{
{\ \lower-1.2pt\vbox{\hbox{\rlap{$<$}\lower5pt\vbox{\hbox{$\sim$}}}}\ }}
\newcommand{\gtrsim}{
{\ \lower-1.2pt\vbox{\hbox{\rlap{$>$}\lower5pt\vbox{\hbox{$\sim$}}}}\ }}
\begin{titlepage}
\begin{center}
{\LARGE\bf High Precision Tests of QED and Physics beyond the Standard Model}\\
\vspace*{0.6cm}

Rafel Escribano$^{1,{\displaystyle \ast}}$,
Eduard Mass\'o$^{1,{\displaystyle \dag}}$\\
\vspace*{0.2cm}

{\footnotesize
$^1$ Grup de F\'\i sica Te\`orica and IFAE, Edifici Cn,
     Universitat Aut\`onoma de Barcelona,
     E-08193 Bellaterra, Spain}\\
\end{center}


\vspace*{1.6cm}

\begin{abstract}
We study the four most significant high precision observables
of QED ---the anomalous electron and muon magnetic moments,
the hydrogen Lamb shift and muonium hyperfine splitting---
in the context of $SU(2) \otimes U(1)$ gauge-invariant effective
Lagrangians. The agreement between the theoretical predictions
for these observables and the experimental data
places bounds on the lowest dimension operators of the
effective Lagrangians. We also place bounds on such effective
operators using other experimental data.
Comparison of the two types of bounds allows us to discuss
the potential of each one of the four high precision observables
in the search for physics beyond the Standard Model.
We find that the anomalous electron and muon magnetic moments are
sensitive to new physics while the hydrogen Lamb shift and muonium
hyperfine splitting are not.
\end{abstract}

\vfill

\footnoterule
{\footnotesize
\noindent ${}^{\displaystyle \ast}$ E-mail: {\tt escribano@ifae.es}\\
\noindent ${}^{\displaystyle \dag}$ E-mail: {\tt masso@ifae.es}}
\end{titlepage}

\section{Introduction and motivation}

QED is the textbook example of the triumph of quantum field theory:
it is a consistent and predictive theory that agrees with experiment to
a very high accuracy
\cite{Kinoshita}.
We know, of course, that QED is a low energy remnant of the more
complete Standard Model (SM) of electroweak interactions. Still, QED is
usually treated as a self-contained theory, into which one may incorporate
the corrections from electroweak and strong interactions. Also, and this
is important for the present article, any non-standard deviations from
QED are assumed to come from extensions of QED that respect the $U(1)$
electromagnetic gauge invariance.

This last point is illustrated by a classical example.
Consider the anomalous magnetic moment of the electron
$a_e \equiv (g - 2)_e/2$.
To parameterise deviations from QED one introduces the $U(1)$ invariant
effective Lagrangian:
\begin{equation}
{\cal L} = \frac{\alpha_1}{\Lambda}\ \overline{\psi_e}\,
\sigma^{\mu\nu}\, \psi_e\, F_{\mu\nu}
\label{leff1}
\end{equation}
Here $\alpha_1$ is a coupling constant and $\Lambda$ is a energy scale.
The (tree-level) contribution to $a_e$ is
\begin{equation}
\delta a_e = 2\,\frac{\alpha_1}{\Lambda}\, \frac{2\, m_e}{e}
\end{equation}
The agreement between the experimental measure
\cite{deltaaeexp}
and the theoretical prediction
\cite{deltaaeth}
for $a_e$ sets the stringent limit
\begin{equation}
-6.9 \times 10^{-11} \leq \delta a_e \leq 4.3 \times 10^{-11}
\label{limitexpdeltaae}
\end{equation}
This limit on $\delta a_e$ (and on all other observables in the article)
is obtained at the 95\% C. L.

In this fashion one can obtain an upper bound on the coefficients of the
effective Lagrangian:
\begin{equation}
\frac{\alpha_1}{\Lambda} \lesssim 2 \times 10^{-5}\ \mbox{TeV}^{-1}
\label{limitalphalambda}
\end{equation}

Although there is nothing wrong with this type of analysis, we think one
can and should go beyond it. One of the reasons is due to the well-known
success of the standard $SU(2)\otimes\,U(1)$ model in describing the
electroweak data. Deviations from the SM have been parameterised in
terms of effective Lagrangians that respect
$SU(2)\otimes\, U(1)$ gauge invariance
\cite{leff}.
Here, we will follow the same prescription, namely we
will describe the effects of physics beyond the SM by a set
of $SU(2)\otimes\, U(1)$ gauge-invariant effective Lagrangians that
modify the high precision QED observables.
In fact, in the  example we have presented, where the Lagrangian
(\ref{leff1}) is used, the allowed values of $\Lambda$ are much greater
than the Fermi scale\footnote{One can estimate \cite{Wudka}
the coupling constant $\alpha_1$ in (\ref{limitalphalambda}) to be of order
$\alpha_1 \approx e/16\pi^2 \simeq 10^{-3}$.}
and thus it should be regarded as natural to use the full $SU(2)\otimes\,
U(1)$ invariance instead of the electromagnetic $U(1)$ invariance.
We further remark that by using the full $SU(2)\otimes\, U(1)$
gauge group we are sensitive to physics beyond the SM rather
than just to QED. Since the SM includes QED we have
widened the scope of the effective Lagrangian approach; going from
the framework where Eq.\ (\ref{leff1}) and (\ref{limitalphalambda})
hold to the analysis performed here.

In this article we will study the four high precision QED observables
that are known with the greatest precision \cite{Kinoshita}:
$(g-2)_e,\, (g-2)_\mu$, the Lamb shift and
muonium hyperfine splitting. Experimental data on such observables
(we call these experiments ``QED experiments'') res\-trict the
coefficients of the lowest-dimension operators in the effective
Lagrangian approach.
As we will see, the operators that lead
to modifications of QED observables will also alter
other quantities measured in other experiments like LEP (we call
these experiments ``non-QED'').
This fact can be used to compare the ability
of different experiments to push the search for new physics.
Both QED and non-QED experiments restrict the coefficients of the
effective operators.
Which experiments lead to the most restrictive limits will tell us
whether, for a particular QED observable, the high precision QED
tests are or are not competitive with non-QED experiments in the
search for physics beyond the Standard Model.
In the article, we will first calculate for each QED observable
the bounds on all the effective operators from QED experiments and
afterwards from non-QED experiments. At the end, we will discuss our
results and compare various bounds. Some of our conclusions may be
relevant in the light of the upcoming experiment
\cite{AGS}
at the Brookhaven Alternative Gradient Synchroton (AGS) to measure the
anomalous magnetic moment of the muon with a precision
$\Delta a_\mu = \pm \, 4 \times 10^{-10}$.

\section{The electron anomalous magnetic moment}

The leading contributions to $a_e$ come from the following
two dimension six operators
\begin{equation}
\label{opeffeB}
{\cal O}_{eB} \equiv
\overline{L_e}\, \sigma^{\mu\nu}\, e_R\, \Phi\, B_{\mu\nu}
\end{equation}
and
\begin{equation}
\label{opeffeW}
{\cal O}_{eW} \equiv
\overline{L_e}\, \sigma^{\mu\nu}\, {\bf \vec \tau}\, e_R\, \Phi\,
{\bf W}_{\mu\nu}
\end{equation}
where $L_e$ is the left-handed isodoublet containing $e_L$,
$e_R$ is its right-handed partner,
${\bf W}_{\mu \nu}$ and $B_{\mu \nu}$ are the
$SU(2)$ and $U(1)$ field strengths, $\Phi$ is the scalar doublet, and
${\bf \vec \tau}$ are the Pauli matrices.

Let us now in turn analyse the effects of these two operators.
The effective Lagrangian corresponding to the operator
(\ref{opeffeB})
is
\begin{equation}
{\cal L} = \frac{\alpha_{eB}}{\Lambda^2}\, {\cal O}_{eB}
\label{leffeB}
\end{equation}
where $\alpha_{eB}$ is a coupling constant and $\Lambda$ is a high energy
scale. After electroweak symmetry breaking, the shift in $a_e$ is
\begin{equation}
\delta a_e = \frac{\sqrt{2}}{v}\, \frac{2\, m_e}{e}\, c_W\,
\epsilon_{eB}
\label{deltaaeeB}
\end{equation}
where $\epsilon_{eB} \equiv \alpha_{eB}\, v^2/ \Lambda^2$
($v \simeq 246\ \mbox{GeV}$ is the Fermi scale).
Hereafter we use $c_W \equiv \mbox{cos}\,\theta_W$
and $s_W \equiv \mbox{sin}\,\theta_W$.
The limit (\ref{limitexpdeltaae}) sets a bound on the
parameter $\epsilon_{eB}$:
\begin{equation}
-5 \times 10^{-6} \leq \epsilon_{eB} (a_e) \leq 3 \times 10^{-6}
\label{limitepsiloneB}
\end{equation}
Here, $a_e$ inside the parentheses
indicates that the limit on $\epsilon_{eB}$ is obtained from the
considera\-tion of the high precision QED observable $a_e$.

The Lagrangian (\ref{leffeB}) also leads to a modification of the
standard $Ze^+e^-$ coupling. The shift in the
$\Gamma_e = \Gamma(Z \longrightarrow e^+e^-)$ width is
\begin{equation}
\frac{\delta \Gamma_e}{\Gamma_e} =
\frac{s_W^2}{g_V^2 + g_A^2}\, \epsilon_{eB}^2
\end{equation}
where $g_V = -1/2 + 2 s_W^2$ and $g_A = -1/2$.
$\Gamma_e$ is measured at the $Z$-peak at LEP
\cite{LEPexp},
and it agrees well with the standard model prediction. One finds
the restriction
\begin{equation}
|\epsilon_{eB} \mbox{\footnotesize (non-QED)}|
\leq 9 \times 10^{-2}
\label{epsiloneBnon-QED}
\end{equation}
where now ``non-QED'' inside the parentheses signifies
that we obtain the limit using
experiments other than high precision QED observations.

The operator (\ref{opeffeW}) also contributes to $a_e$.
Writing
\begin{equation}
{\cal L} =  \frac{\alpha_{eW}}{\Lambda^2}\, {\cal O}_{eW}
\label{leffeW}
\end{equation}
we find
\begin{equation}
\delta a_e = -\frac{\sqrt{2}}{v}\, \frac{2\, m_e}{e}\, s_W\,
\epsilon_{eW}
\label{deltaaeeW}
\end{equation}
with $\epsilon_{eW} \equiv \alpha_{eW}\, v^2/ \Lambda^2$.
Using (\ref{limitexpdeltaae}), we get
\begin{equation}
-5 \times 10^{-6}\leq \epsilon_{eW} (a_e) \leq 8 \times 10^{-6}
\label{limitepsiloneW}
\end{equation}

The operator ${\cal O}_{eW}$ leads to couplings $Ze^+e^-$ and
$We\nu$ that would modify the standard model predictions.
We find, however, that the possible shift in
$Z \longrightarrow e^+e^-$ decay leads to the most restrictive limits of
all the ``non-QED'' experiments.
We obtain
\begin{equation}
|\epsilon_{eW} \mbox{\footnotesize (non-QED)}|
\leq 5 \times 10^{-2}
\label{epsiloneWnon-QED}
\end{equation}

We should now comment on
the question of cancellations among different effective
contributions. The effective Lagrangian is a linear combination of both
operators in (\ref{opeffeB}) and (\ref{opeffeW}), and the total contribution
to $a_e$ is the sum of both contributions in
(\ref{deltaaeeB}) and (\ref{deltaaeeW}).
A strong cancellation in the two contributions either to $a_e$
(or to $\Gamma_e$) would be unnatural. Still, a partial cancellation
could occur and thus the limits could be relaxed but presumably only
by a factor of order one. Fortunately, our main conclusions depend
only on the order of magnitude of the limit and not on such details.
Consequently, we will assume that there are no fine-tuned
cancellations among contributions to the observables.

\section{The muon anomalous magnetic moment}

There are two operators, similar to (\ref{opeffeB}) and (\ref{opeffeW}),
that contribute to $a_\mu$:
\begin{equation}
\begin{array}{rcl}
{\cal O}_{\mu B}&\equiv&
\overline{L_\mu}\, \sigma^{\mu\nu}\, \mu_R\, \Phi\, B_{\mu\nu}\\[1ex]
{\cal O}_{\mu W}&\equiv&
\overline{L_\mu}\, \sigma^{\mu\nu}\, {\bf \vec \tau}\, \mu_R\,
\Phi\, {\bf W}_{\mu\nu}
\label{opeffmu}
\end{array}
\end{equation}

The analysis is very similar to the case of $a_e$. The agreement
between theory and experiment
\cite{deltaamuth,deltaamuexp}
restricts any contribution to $a_\mu$ as follows:
\begin{equation}
-1.4 \times 10^{-8}\leq \delta a_\mu \leq 2.2 \times 10^{-8}
\end{equation}
which implies
\begin{equation}
\begin{array}{rcl}
-4 \times 10^{-6}\leq &\epsilon_{\mu B}(a_\mu)& \leq
7 \times 10^{-6}\\[1ex]
-2 \times 10^{-5}\leq &\epsilon_{\mu W}(a_\mu)& \leq
7 \times 10^{-6}
\end{array}
\label{limitepsilonmu}
\end{equation}
(The parameters $\epsilon_{\mu B}$ and $\epsilon_{\mu W}$ are defined
in analogy to $\epsilon_{eB}$ and $\epsilon_{eW}$).

The operators (\ref{opeffmu}) modify
$\Gamma(Z \longrightarrow \mu^+\mu^-)$. The LEP data imply
\begin{eqnarray}
|\epsilon_{\mu B} \mbox{\footnotesize (non-QED)}|
&\leq& 9 \times 10^{-2}
\label{epsilonmuBnon-QED}\\
|\epsilon_{\mu W} \mbox{\footnotesize (non-QED)}|
&\leq& 5 \times 10^{-2}
\label{epsilonmuWnon-QED}
\end{eqnarray}
${\cal O}_{\mu W}$ contains vertices like $W\mu\nu$ that modify
for instance $\mu \longrightarrow e\nu\nu$. However,
the corres\-ponding limit on $\epsilon_{\mu W}$ is much less stringent
than (\ref{epsilonmuWnon-QED}).

\section{The Lamb shift}

The splitting of the hydrogen levels $2S_{1/2}$ and $2P_{1/2}$,
$\Delta E_H (2S_{1/2}-2P_{1/2}) \equiv E_{\mbox{\scriptsize LS}}$,
known as the
Lamb shift, is an important observable to test QED. The agreement between
experiment
\cite{Lambexp}
and theory
\cite{Lambth}
requires that other contributions to
the Lamb shift respect the stringent limit
\begin{equation}
-38 \leq \delta E_{\mbox{\scriptsize LS}} \leq 10\ \mbox{kHz}
\label{limitLS}
\end{equation}

There is a long list of dimension six operators that could contribute to
the Lamb shift. However, after discarding the effective operators that induce
redefinitions of the physical parameters and using the equations of motion
in a rigorous way, one can select the following independent basis
\cite{R&Ework_in_preparation}:
\begin{equation}
\left\{{\cal O}_{eB},{\cal O}_{eW},{\cal O}_{\partial B},{\cal O}_{DW}
\right\}
\end{equation}
where ${\cal O}_{eB}$, ${\cal O}_{eW}$ are defined in (\ref{opeffeB})
and (\ref{opeffeW}), and
\begin{equation}
\begin{array}{rcl}
{\cal O}_{\partial B}&\equiv&
\partial_\lambda B^{\mu\nu}\, \partial^\lambda B_{\mu\nu}\\[1ex]
{\cal O}_{DW}&\equiv&
\left[D_\lambda {\bf W}^{\mu\nu}\right]^{\dag}
\left[D^\lambda {\bf W}_{\mu\nu}\right]
\label{opeffpartialD}
\end{array}
\end{equation}

Let us start with the first operator, ${\cal O}_{eB}$. Its effects are
expressed via the Lagrangian (\ref{leffeB}), that arose earlier.
Its contribution to the Lamb shift is given by
\begin{equation}
\delta E_{\mbox{\scriptsize LS}} = \frac{(m_e\, \alpha)^3}{6\, \pi}\,
\frac{e}{2\, m_e}\,
\frac{\sqrt{2}}{v}\, c_W\, \epsilon_{eB}
\label{deltaELSeW}
\end{equation}
The experimental limit (\ref{limitLS}) leads to
\begin{equation}
-7 \times 10^{-3} \leq \epsilon_{eB} (E_{\mbox{\scriptsize LS}})
\leq 2 \times 10^{-3}
\end{equation}
The Lagrangian (\ref{leffeW}), containing ${\cal O}_{eW}$, has a
contribution similar to (\ref{deltaELSeW}),
with $c_W\, \epsilon_{eB} \rightarrow -s_W\, \epsilon_{eW}$.
The corresponding restriction is
\begin{equation}
-3 \times 10^{-3}\leq \epsilon_{eW} (E_{\mbox{\scriptsize LS}})
\leq 2 \times 10^{-2}
\end{equation}
where $E_{\mbox{\scriptsize LS}}$ inside the parentheses indicates
that the limit is obtained using $E_{\mbox{\scriptsize LS}}$.

While these two operators affect the $ee\gamma$ vertex, the operators
(\ref{opeffpartialD}) contribute to the Lamb shift
through the photon self-energy. We find
\begin{equation}
\delta E_{\mbox{\scriptsize LS}} = m_e\, \alpha^4\, \frac{m_e^2}{v^2}\,
\left(c_W^2\, \epsilon_{\partial B} + s_W^2\, \epsilon_{DW}\right)
\end{equation}
where $\epsilon_{\partial B}$ and $\epsilon_{DW}$ are defined in analogy to
$\epsilon_{eB}$ and $\epsilon_{eW}$. Assuming that there are no
cancellations among the contributions of
${\cal O}_{\partial B}$ and ${\cal O}_{DW}$, yields
\begin{equation}
\begin{array}{rcl}
-6 \times 10^{3} \leq &\epsilon_{\partial B} (E_{\mbox{\scriptsize LS}})&
\leq 2 \times 10^{3}\\[1ex]
-2 \times 10^{4} \leq &\epsilon_{DW} (E_{\mbox{\scriptsize LS}})&
\leq 5 \times 10^{3}
\label{limitepsilonpartialD}
\end{array}
\end{equation}

Following our general strategy we now calculate the limits to the different
$\epsilon$'s using other experimental data. The limits on $\epsilon_{eB}$
and $\epsilon_{eW}$ have already been quoted in (\ref{epsiloneBnon-QED})
and (\ref{epsiloneWnon-QED}). The best bounds on $\epsilon_{\partial B}$
and $\epsilon_{DW}$ come from the LEP measurements on $Z$ widths. They are
\begin{equation}
\begin{array}{rcl}
-2 \times 10^{-2} \leq &\epsilon_{\partial B} \mbox{\footnotesize (non-QED)}&
\leq 2 \times 10^{-2}\\[1ex]
-6 \times 10^{-3}\leq &\epsilon_{DW} \mbox{\footnotesize (non-QED)}&
\leq 4 \times 10^{-3}
\end{array}
\end{equation}

\section{Muonium hyperfine splitting}

Muonium is a system which displays many of the hydrogen properties but does
not contain constituent hadrons. It is in this respect a good testing ground
for QED. Its ground state hyperfine splitting,
$\nu_{\mu \mbox{\scriptsize -hfs}}$,
corresponds to the energy difference among states with parallel or
antiparallel alignment of the $e^-$ and $\mu^+$ magnetic moments.
It has been measured very accurately
\cite{muoniumexp}
and there are precise theoretical calculations
\cite{muoniumth}.
Additional contributions to this observable are limited by
\begin{equation}
-2.5 \leq \delta \nu_{\mu \mbox{\scriptsize -hfs}} \leq 3.0\ \mbox{kHz}
\label{limitmuonium}
\end{equation}

The independent dimension six operators contributing to
$\nu_{\mu \mbox{\scriptsize -hfs}}$ can be classified into two types. We have,
first,
${\cal O}_{e B}$ and ${\cal O}_{e W}$ that affect the $ee\gamma$ vertex and
${\cal O}_{\mu B}$ and ${\cal O}_{\mu W}$ that affect the $\mu \mu \gamma$
vertex. These four operators have already appeared in our analysis.
The second type are four-fermion operators. Using Fierz shuffling,
one can select the following complete set of effective operators that we call
${\cal O}_{4f}$:
\begin{equation}
{\cal O}_{4f} = \left\{{\cal O}_{\ell \ell}^{(1)},
{\cal O}_{\ell \ell}^{(3)}, {\cal O}_{e \mu}, {\cal O}_{\ell \mu},
{\cal O}_{e \ell}\right\}
\end{equation}
where
\begin{equation}
\label{opeff4f}
\begin{array}{rcl}
{\cal O}_{\ell \ell}^{(1)}&\equiv&
(\overline {L_e}\, \gamma^\mu\, {L_e})
(\overline {L_\mu}\, \gamma_\mu\, {L_\mu})\\[1ex]
{\cal O}_{\ell \ell}^{(3)}&\equiv&
(\overline {L_e}\, \gamma^\mu\, {\bf \vec \tau}\, {L_e})
(\overline {L_\mu}\, \gamma_\mu\, {\bf \vec \tau}\, {L_\mu})\\[1ex]
{\cal O}_{e \mu}&\equiv&
(\overline {e_R}\, \gamma^\mu\, e_R)
(\overline {\mu_R}\, \gamma_\mu\, \, \mu_R)\\[1ex]
{\cal O}_{\ell \mu}&\equiv&
(\overline {L_e}\, \gamma^\mu\, {L_e})
(\overline {\mu_R}\, \gamma_\mu\, \, \mu_R)\\[1ex]
{\cal O}_{e \ell}&\equiv&
(\overline {e_R}\, \gamma^\mu\, e_R)
(\overline {L_\mu}\, \gamma_\mu\, {L_\mu})
\end{array}
\end{equation}

The contribution of ${\cal O}_{e B}$ and ${\cal O}_{e W}$ is calculated to be
\begin{equation}
\delta \nu_{\mu \mbox{\scriptsize -hfs}} =
\frac{8}{3\, \pi}\, \alpha^2\, R_\infty\, \frac{m_e}{m_\mu}\,
\left[\frac{2 m_e}{e}\, \frac{\sqrt{2}}{v}\,
(c_W\, \epsilon_{eB}-s_W\, \epsilon_{eW})\right]
\label{deltanumuoniumeBW}
\end{equation}
and that from ${\cal O}_{\mu B}$ and ${\cal O}_{\mu W}$ is similar to
(\ref{deltanumuoniumeBW}), with $m_e \rightarrow m_\mu$ inside the brackets.
The four-fermion effective operators contribute as
\begin{equation}
\delta \nu_{\mu \mbox{\scriptsize -hfs}} =
\frac{\alpha}{\pi^2}\, R_\infty\, \frac{m_e^2}{v^2}\,
\left(\epsilon_{\ell \ell}^{(1)}+\epsilon_{\ell \ell}^{(3)}+
      \epsilon_{e \mu}-\epsilon_{\ell \mu}-\epsilon_{e \ell}
\right)
\end{equation}
where $R_\infty$ is the Rydberg constant.

Again excluding fortuitous cancellations,
we use (\ref{limitmuonium}) to find
\begin{equation}
\begin{array}{rcl}
-4 \times 10^{-2} \leq &\epsilon_{eB}(\nu_{\mu \mbox{\scriptsize -hfs}})&
\leq 4 \times 10^{-2}\\[1ex]
-8 \times 10^{-2} \leq &\epsilon_{eW}(\nu_{\mu \mbox{\scriptsize -hfs}})&
\leq 6 \times 10^{-2}\\[1ex]
-2 \times 10^{-4} \leq &\epsilon_{\mu B}(\nu_{\mu \mbox{\scriptsize -hfs}})&
\leq 2 \times 10^{-4}\\[1ex]
-4 \times 10^{-4} \leq &\epsilon_{\mu W}(\nu_{\mu \mbox{\scriptsize -hfs}})&
\leq 3 \times 10^{-4}
\end{array}
\end{equation}
and
\begin{equation}
\label{limitepsilon4fQED}
\begin{array}{rcl}
-40 \leq &\epsilon_{\ell \ell}^{(1)}(\nu_{\mu \mbox{\scriptsize -hfs}})& \leq
50\\[1ex]
-40 \leq &\epsilon_{\ell \ell}^{(3)}(\nu_{\mu \mbox{\scriptsize -hfs}})& \leq
50\\[1ex]
-40 \leq &\epsilon_{e \mu}(\nu_{\mu \mbox{\scriptsize -hfs}})& \leq 50\\[1ex]
-50 \leq &\epsilon_{\ell \mu}(\nu_{\mu \mbox{\scriptsize -hfs}})& \leq
40\\[1ex]
-50 \leq &\epsilon_{e \ell}(\nu_{\mu \mbox{\scriptsize -hfs}})& \leq 40\\[1ex]
\end{array}
\end{equation}
where we have defined $\epsilon_i \equiv \alpha_i\, v^2/ \Lambda^2$
and the $\alpha_i$'s are the corresponding coefficients of the operators
(\ref{opeffeB}), (\ref{opeffeW}), (\ref{opeffmu}),
and (\ref{opeff4f}) in the effective Lagrangian.
Now $\nu_{\mu \mbox{\scriptsize -hfs}}$ inside the parentheses indicates that
the
limit is obtained using $\nu_{\mu \mbox{\scriptsize -hfs}}$.

As before, we now use other experimental data to limit the
$\epsilon$ parameters. The best limits on $\epsilon_{eB}, \epsilon_{eW},
\epsilon_{\mu B}$, and $\epsilon_{\mu W}$ are extracted from the LEP data
and have been already quoted in (\ref{epsiloneBnon-QED}),
(\ref{epsiloneWnon-QED}), (\ref{epsilonmuBnon-QED}), and
(\ref{epsilonmuWnon-QED}).

The new operators (\ref{opeff4f})
have vertices that modify the standard prediction for
$e^+e^- \longrightarrow \mu^+ \mu^-$ and ${\cal O}_{\ell \ell}^{(3)}$ also
modifies the $Z$-widths. LEP data restrict all these operators.
Additionally, the operator ${\cal O}_{\ell \ell}^{(3)}$
alters the $\mu$-decay prediction but the restriction is less severe.
Finally, we obtain
\begin{equation}
\begin{array}{rcl}
&|\epsilon_{\ell \ell}^{(1)}(\mbox{\footnotesize non-QED})|&
\leq 9 \times 10^{-1}\\[1ex]
-3 \times 10^{-3} \leq
&\epsilon_{\ell \ell}^{(3)}(\mbox{\footnotesize non-QED})&
\leq 2 \times 10^{-3}\\[1ex]
&|\epsilon_{e \mu}(\mbox{\footnotesize non-QED})|&
\leq 9 \times 10^{-1}\\[1ex]
&|\epsilon_{\ell \mu}(\mbox{\footnotesize non-QED})|&
\leq 9 \times 10^{-1}\\[1ex]
&|\epsilon_{e \ell}(\mbox{\footnotesize non-QED})|&
\leq 9 \times 10^{-1}
\end{array}
\end{equation}

\section{Summary and discussion}

We have studied four high precision observables that provide
excellent tests of QED. For each observable we have identified the effective
Lagrangians that can contribute to it.
The Lagrangians are composed of the (lowest-dimension) independent operators
that are $SU(2)\otimes\, U(1)$ gauge-invariant.

There are two steps in our calculations. We have, first, bounded all
the $\epsilon$ coefficients of the effective Lagrangian using
QED experiments and theoretical predictions.
The best bounds are given in
(\ref{limitepsiloneB}), (\ref{limitepsiloneW}),
(\ref{limitepsilonmu}), (\ref{limitepsilonpartialD}), and
(\ref{limitepsilon4fQED}).
Second, we have noticed that the operators in the effective Lagrangian
contain terms that lead to new effects in observables other than the
above four.
We can thus use further data to bound the same coefficients,
but now the data is not from the ``QED'' observables but rather from
``non-QED'' observables. In fact, the most restrictive ``non-QED'' data
turns out to be LEP data.

We have bounds from two experimental sources. As we said in the
introduction a compari\-son between them is enlightening since it
is clear that the experiment placing the strongest bounds on the effective
coefficients $\epsilon$'s is the one most sensitive to new physics.
For a given $\epsilon$, the comparison is done in the effective
Lagrangian approach. Thus, our conclusions are expected to be model
independent.

The anomalous magnetic moments of the electron and the muon restrict the
coefficients of the operators ${\cal O}_{eB}, {\cal O}_{eW}$, and
${\cal O}_{\mu B}, {\cal O}_{\mu W}$ much more severely that any ``non-QED''
data. As a first conclusion,
this suggests that by improving the high precision measurements of
$(g-2)_e$ and $(g-2)_\mu$ one could be sensitive to physics beyond the
standard electroweak model. In the light of this remark, we think it is
interesting that $(g-2)_\mu$ will be measured with unprecedent precision
at the AGS
\cite{AGS}.

The conclusion we reach for the other two observables, the hydrogen
Lamb shift and the muonium hyperfine splitting, is the opposite.
Looking at the numerical limits obtained in this article, we see that the
limits on the operators
${\cal O}_{eB}, {\cal O}_{eW}, {\cal O}_{\mu B}$, and ${\cal O}_{\mu W}$
obtained from the hydrogen Lamb shift and muonium hyperfine
splitting are weaker than the limits using $(g-2)_e$ and $(g-2)_\mu$.
For the remaining operators, namely
${\cal O}_{\partial B}, {\cal O}_{DW}$, and ${\cal O}_{4f}$,
bounds from LEP data are more stringent.
This suggests than these two observables are far from being sensitive
to new physics.
\vspace*{0.6cm}

\noindent
{\it Note added:}
After we finished the writing, we became aware of a related work published
in \cite{Perrottet}. In this reference, the authors have computed non-standard
contributions to $a_\mu$ arising from composite fermions and gauge bosons,
and have compared with constraints from LEP-2 when available. Our work
differs from theirs in the following aspects. We use gauge-invariant
effective Lagrangians and calculate at tree-level, while they calculate loops
with form factors, excited leptons, etc. Also, they do a very exhaustive
study but restricted to $a_\mu$ while we have included all the relevant
QED observables.
\vspace*{0.6cm}

\noindent
{\it Acknowledgments:}
We acknowledge financial support from the CICYT AEN\-95-0815 project and
from the Theo\-retical Astroparticle Network under EEC
Contract No.~CHRX-CT93-0120 (Direction Gene\-rale 12 COMA).
We thank M.~J.~Lavelle for a critical reading of the manuscript. One of us
(R.~E.) would like to thank A.~Bramon for his help and many fruitful
discussions, and M.~Mart\'{\i}nez and F.~Teubert for their comments and
helpful collaboration in providing the most recent LEP-I data.
R.~E.~also thanks the CERN Theory Division for their warm hospitality and
acknowledges financial support by an F.~P.~I.~grant from the Universitat
Aut\`onoma de Barcelona.

\end{document}